\documentclass[aps,pra,preprint,floatfix,groupedaddress,showkeys]{revtex4-1}

\usepackage{graphicx}
\usepackage{dcolumn}
\usepackage{amsmath,bm}
\usepackage{color}%

\begin{document}


\title{Rigid Unit Modes in $sp$-$sp^2$ Hybridized Carbon Systems:
Origin of Negative Thermal Expansion}

\author{Cheol-Woon Kim}
\affiliation{Department of Physics and
             Research Institute for Basic Sciences,
             Kyung Hee University, Seoul, 02447, Korea}

\author{Seoung-Hun Kang}
\affiliation{Department of Physics and
             Research Institute for Basic Sciences,
             Kyung Hee University, Seoul, 02447, Korea}
             
\author{Young-Kyun Kwon}
\email[Corresponding author. E-mail: ]{ykkwon@khu.ac.kr}
\affiliation{Department of Physics and
             Research Institute for Basic Sciences,
             Kyung Hee University, Seoul, 02447, Korea}

\date{\today}

\begin{abstract}
Using density functional theory combined with quasi-harmonic
approximation, we investigate the thermal expansion behaviors of 
three different types ($\alpha$, $\beta$, and $\gamma$) of graphyne
which is a two-dimensional carbon allotrope composed of $sp$ and
$sp^2$ bonds. For each type of graphyne, we obtain the temperature
dependent area variation by minimizing its free energy as a function
of temperature, which is calculated by considering all the phonon
modes in the whole Brillouin zone. We find that all three types of
graphyne exhibit negative in-plane thermal expansion up to
$T\lesssim1000$~K. The observed in-plane thermal contraction can be
attributed partially to the ripple effect, similarly in graphene.
The ripple effect itself, however, is not sufficient to explain
anomalously larger thermal contraction found in graphyne than in
graphene. Our deliberate analysis on the phonon modes observed in
graphyne enables us to discover another source causing such thermal
expansion anomaly. We find that there are particular phonon modes with
frequencies around a few hundreds of cm$^{-1}$ existing exclusively in
graphyne that may fill empty spaces resulting in area reduction. These
modes are identified as ``rigid unit modes'' corresponding to the
libration of each rigid unit composed of $sp^2$ bonds.
\end{abstract}



\maketitle


\section{Introduction}
\label{Introduction}

Various carbon allotropes based on $sp^2$ hybridization, such as
fullerenes,~\cite{Kroto1985} nanotubes,~\cite{Iijima1991} and 
graphene,~\cite{Novoselov2004} have been intensively studied during
last a few decades, since their first discoveries. Their unique
properties are basically related to the hexagonal geometry and the
Dirac cones in the electronic structure of
graphene,~\cite{Zhang2005,Novoselov2005,CastroNeto2009} which is a
base material to form other $sp^2$-bonded carbon allotropes. To make
use of graphene in future nanoelectronics, however, it is
indispensable to open an energy gap at the Fermi level $E_F$. Although
lots of attempts have been made to open a band gap,~\cite{{Sahin2011},
{Kim2008},{Son2006},{Zhou2007},{Kwon2010},{Haskins2011}} nothing has
yet come along for practical applications. As a separate approach,
sundry efforts have been made to search for other 2D allotropes
intrinsically possessing an energy gap at $E_F$ to be used as graphene
substitutes in electronic applications.~\cite{Hirsch2010}

\begin{figure}[t]
\includegraphics[width=1.0\columnwidth]{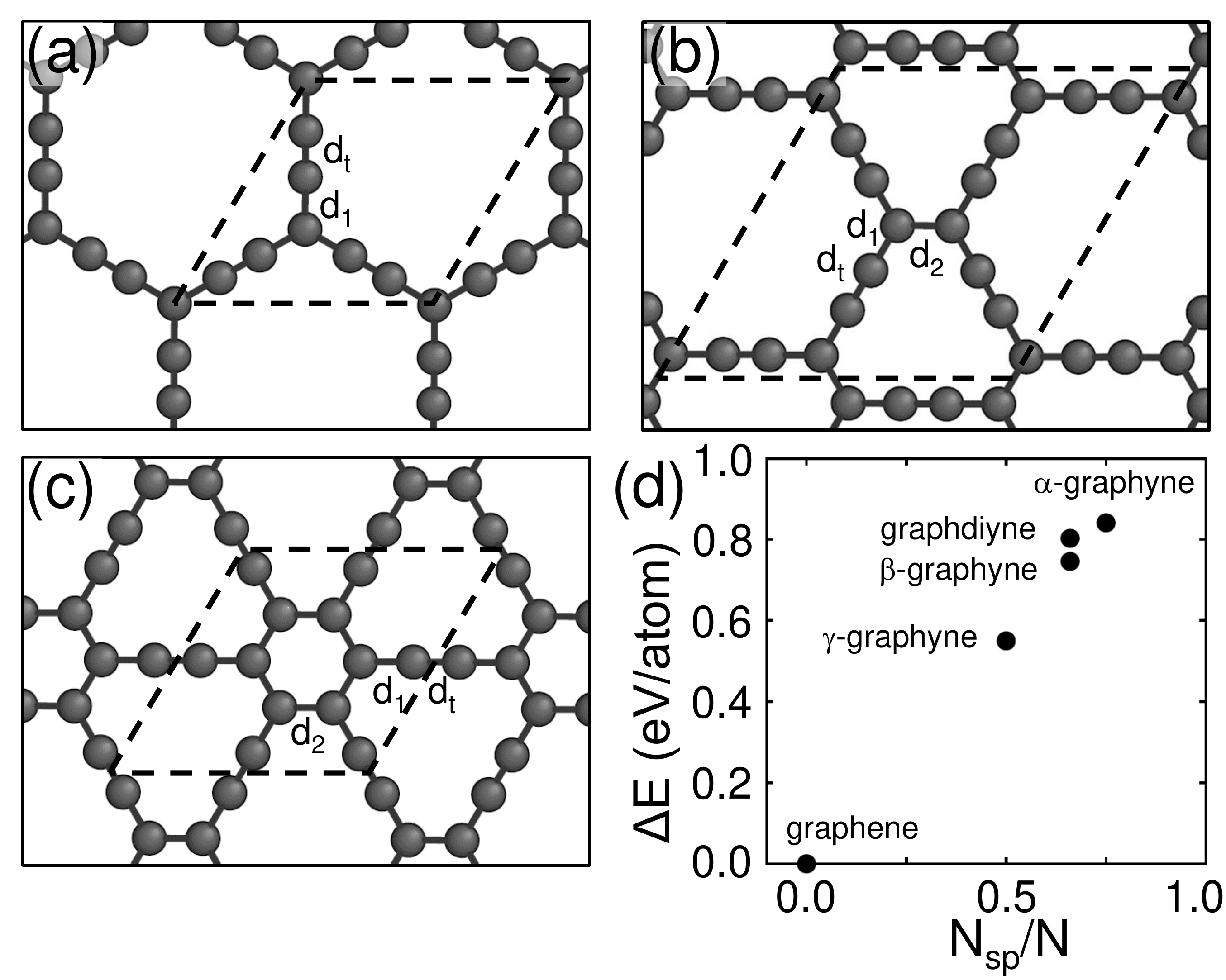}
\caption{Equilibrium structures of (a) $\alpha$-graphyne, (b)
$\beta$-graphyne and (c) $\gamma$-graphyne. Respective hexagonal unit
cells are depicted by dashed parallelograms. Nonequivalent bonds in
each type of graphyne are denoted by $d_1$, $d_2$, and $d_t$. $d_1$
and $d_2$, which are single-bond like, represent a bond between an
$sp^2$-bonded (or triply-coordinated) and an $sp$-bonded (or
doubly-coordinated) atoms, and one between two triply-coordinated
atoms, respectively, whereas $d_t$ designates a triple bond between
two $sp$ atoms in every acetylenic linkage. (d) Cohesive energy
comparison of three types of graphyne and graphdiyne with respect to
graphene as a function of $N_{sp}/N$, where $N_{sp}$ and $N$ are the
number of $sp$-bonded or doubly-coordinated atoms and $N$ and the
total number of atoms in the unit cell, respectively.
\label{structure}}
\end{figure}

A new 2D layered carbon allotrope called graphyne was proposed in
1987.~\cite{Baughman1987} It can be geometrically generated from
graphene by inserting a single acetylenic linkage (AL) consisting of
two $sp$-hybridized carbon atoms into a bond in graphene composed of
$sp^2$-hybridized carbon atoms only. Since there are a number of ways
of inserting ALs, many different types of graphyne can be formed.
Among them were mainly considered three types with a hexagonal
symmetry, classified by the number of ALs attached to each
$sp^2$-hybridized or triply-coordinated carbon atom: $\alpha$-,
$\beta$-, and $\gamma$-graphyne. Fig.~\ref{structure}(a--c) shows
their equilibrium structures. In $\alpha$-graphyne
shown in Fig.~\ref{structure}(a), there are three ALs connected to
each $sp^2$-hybridized carbon atom, whereas in $\beta$-graphyne
($\gamma$-graphyne), every $sp^2$-bonded atom has only two ALs (one
AL) as shown in Fig.~\ref{structure}(b) and (c), respectively.

Theoretical studies reported that both $\alpha$- and $\beta$-graphyne
structures possess Dirac cones near $E_F$ in its Brillouin zone (BZ)
representing their semi-metallic characteristics like in
graphene.~\cite{Kang2011,Zhou2011,Malko2012,Kim2012,Liu2012,Popov2013,
Perkgoz2014} It was also reported that $\gamma$-graphyne, which is by
far the most stable graphyne, is interestingly an intrinsic
semiconductor with a direct band gap of 0.46~eV at $M$ points. The
existence of the band gap was ascribed by the Peierls instability
resulting in the wave function localization at the triple
bonds.~\cite{Kim2012} It is, therefore, expected that
$\gamma$-graphyne could be made use of as a semiconducting component
in future nanoelectronics.

Although various research groups have endeavored to synthesize them,
graphyne has not yet been realized in a crystalline form.
Nevertheless, it was already reported that graphyne sub units that
resemble $\gamma$-graphyne were successfully
synthesized.~\cite{{Spitler2006},{Haley2008},{Takeda2010}}
Moreover, one of their cousin structures with two ALs in a sequence
instead of a single AL in $\gamma$-graphyne, called graphdiyne, was
also reported to be discovered in the form of sheet,~\cite{Li2010} as
well as in a tube form.~\cite{Qian2012} According to our calculation
as well as other studies,~\cite{Ivanovskii2013,Yin2013,Shin2014}
$\beta$-graphyne and $\gamma$-graphyne are even more stable than
graphdiyne, as shown in Fig.~\ref{structure}(d), and thus expected to
be synthesized in near future.

Compared to the structural and electronic properties of $\alpha$-,
$\beta$-, and $\gamma$-graphyne that have been intensively studied as
discussed above, there have been not many studies on their thermal and
thermal expansion (TE) behaviors,~\cite{Shao2012,Popov2013,
Perkgoz2014} which are also necessary properties to be disclosed in
advance for them to be utilized as various devices in future
nanoelectronics.

It is well known that most materials expand thermally with a very 
different material-dependent linear TE coefficient (TEC) of
$10^{-7}\sim10^{-4}$ mainly due to the atomic vibration under
asymmetric potential. In almost all devices and equipment, such a
TE usually becomes a critical problem deteriorating their performance
and lifetime. There have been, on the other hand, several reports
revealing materials with a negative TEC (NTEC), such as
graphite,~\cite{Nelson1945} carbon nanotubes,~\cite{Kwon2004,Kwon2005}
zirconium tungstate (ZrW$_2$O$_8$),~\cite{Martinek1968} perovskite
oxides,~\cite{Long2009,Long2010,Azuma2011} and Invar
alloys,~\cite{Moriya1993,Sumiyama1979} originating due to a ripple
effect, bending/twisting modes, rigid unit modes
(RUMs),~\cite{Hammonds1996, Tucker2005} atomic radius
contraction,~\cite{Arvanitidis2003} or a magnetovolume
effects.~\cite{Guillaume1897} Researchers have made various attempts
to make composites containing NTE materials, such as
ZrW$_2$O$_8$~\cite{Sullivan2005,Verdon1997,Holzer2011} and
$\beta$-eucryptite~\cite{Xue2010}, as a TE compensator, expecting them
to be high-performance composites with near zero TE. It was indeed
shown that TE can be suppressed in molten aluminum alloy with NTE
manganese antiperovskites.~\cite{Ishii2008}

In this letter, we report a first discovery of RUMs in two-dimensional
graphitic carbon systems, which are responsible for a TE anomaly of
NTE in $sp$-$sp^2$ hybridized carbon systems, similarly observed in 2D
perovskites.~\cite{Goodwin2005} The thermal properties and TE
behaviors of $\alpha$-, $\beta$-, and $\gamma$-graphyne were
investigated by first-principles calculations of temperature- and
volume-dependent free energies from the phonon properties. Similar to
graphene, all of three configurations contract with temperature up to
$T{\approx}1000$~K, but the absolute values of their TECs are much
larger than that of graphene especially at low temperatures. Based on
our thorough investigation on their phonon modes, we attributed such
thermal contraction behaviors to not only the ripple effect as in
graphene, but also RUMs corresponding to libration motions of rigid
units composed of $sp^2$-bonds only.

\section{Computational details}
\label{Computational}

To investigate the TE behaviors of three types of graphyne, we first
obtained their equilibrium structures by carrying out first-principles
calculations within the density functional theory
(DFT),~\cite{Kohn1965} as implemented in Vienna \textit{ab initio}
simulation package (VASP),~\cite{{Kresse1996},{Kresse1993}} while
adjusting three different carbon-carbon bonds, and thus the lattice
constant of each graphyne as discussed in Sect.~\ref{Results}.
Projector augmented wave potentials~\cite{{Blochl1994},{Kresse1999}}
was employed to describe the valence electrons, and the electronic
wave functions were expanded by a plane wave basis set with the cutoff
energy of 450~eV. The exchange-correlation functional was treated with
the Perdew-Burke-Ernzerhof (PBE) parameterization~\cite{Perdew1996} of
the generalized gradient approximation (GGA). To mimic single-layered
graphyne, we introduced a vacuum region with 15~{\AA} along the
$c$-axis perpendicular to the sheet. The BZ was sampled using a
$\Gamma$-centered $10{\times}10{\times}1$ $k$-grid for the primitive
unit cell of each type of graphyne. Although the primitive cell size
of $\beta$-graphyne is larger than the other types, the same $k$-grid
was employed for precise calculation for $\beta$-graphyne because
there were earlier studies on $\beta$-graphyne showing inconsistent
results on structural stability.~\cite{Popov2013,Perkgoz2014}

The phonon dispersion relations were computed by applying the finite
displacement method (FD)~\cite{Parlinski1997,Togo2008} for
$3{\times}3{\times}1$ supercells. The corresponding reduced BZs were
sampled by $3{\times}3{\times}1$ $k$-point meshes. In each supercell,
we solved the secular equation of dynamical matrix constructed at
every wave vector $\mathbf{q}$ from the force constant matrices
computed under FD to obtain the phonon dispersion relation. To
investigate thermal properties, we employed quasi-harmonic
approximation (QHA),~\cite{Pavone1993,Togo2010} in which the Helmholtz
free energy of a 2D graphyne sheet was calculated by 
\begin{widetext}
\begin{equation}
F(T,A)=U(A)+\frac{1}{2}\sum_{\mathbf{q},n}\hbar\omega_{\mathbf{q},n}
   +k_BT\sum_{\mathbf{q},n}\ln\left\{1
        -\exp\left(-\frac{\hbar\omega_{\mathbf{q},n}}{k_BT}\right)
      \right\},
\label{equation1}
\end{equation}
\end{widetext}
where $\hbar$, and $k_B$ are the reduced Planck constant, and the
Boltzmann constant, respectively; and $\omega_{\mathbf{q},n}$ is the
phonon frequency with the wave vectors $\mathbf{q}$ and the mode
indices $n$. In Eq.~(\ref{equation1}), the first term $U(A)$ is the
system internal energy with a constant area $A$ at $T=0$; the second
term represents the vibrational zero-point energy of the lattice; and
the last term corresponds to the phonon contribution to the free
energy in QHA.~\cite{Pavone1993,Togo2010} $U(A)$ and
$\omega_{\mathbf{q},n}$ were evaluated at 12 different area points.
$F(T,A)$ was fitted to the integral form of the Vinet equation of
state~\cite{Vinet1989} to obtain the minimum values of the
thermodynamic functions with respect to the area, and thus the
equilibrium area, $A(T)$, as a function of temperature. TE behaviors
of three types of graphyne sheets were explored by calculating the
area TEC $\alpha_S(T)$ as a function of $T$,
\begin{equation}
{\alpha}_{S}(T)=\frac{1}{A(T)}\frac{dA(T)}{dT}.
\label{equation2}
\end{equation}

\section{Results and discussion
\label{Results}}

We first searched for the equilibrium structures of three kinds of
graphyne, which are indispensable to their phonon dispersion relations
and thermal properties. Graphyne can be made by replacing all or some
bonds in graphene by an AL (-C$\equiv$C-). Among numerous different
types of graphyne produced in this way, we only focused on three types
of graphyne, $\alpha$, $\beta$, and $\gamma$, which are highly
symmetric in a hexagonal lattice relative to the other types. The
$\alpha$-graphyne can be made by a complete substitution for graphitic
bonds, while the $\beta$-graphyne and $\gamma$-graphyne can be formed
by two-thirds and one-third substitutions, respectively. In these
types of graphyne, there are at most three nonequivalent carbon-carbon
bonds, $d_1$, $d_2$, and $d_t$. Single-bond like $d_1$ and $d_2$
represent bonds from a $sp^2$ atom to a neighboring $sp$ and $sp^2$
atom, respectively, while a triple-bond $d_t$ describes a bond between
two neighboring $sp$ atoms forming an AL together with two $d_1$
bonds.

For each graphyne, we completely scanned the whole energy surface
while adjusting the bond length of three nonequivalent carbon-carbon
bonds, $d_1$, $d_2$, and $d_t$ keeping its hexagonal symmetry. Note
that there are no $d_2$ bonds in $\alpha$-graphyne, and the lattice
constant of each unit cell is uniquely determined for each set of
given $d_1$, $d_2$, and $d_t$. The equilibrium structures of three
types of graphyne were determined at the minimum energy point in each
energy surface, and shown in Fig.~\ref{structure}(a--c), where their
hexagonal lattices were depicted by dashed parallelograms representing
their unit cells. Two nonequivalent bonds in the $\alpha$-graphyne
were calculated to be $d_1=1.397$~{\AA} and $d_t=1.230$~{\AA} with
lattice constant of $a_{\mathrm{eq}}=6.969$~{\AA}. For the
$\beta$-graphyne ($\gamma$-graphyne), on the other hand, its
calculated lattice constant and three nonequivalent bond lengths are
$a_{\mathrm{eq}}=9.480$~{\AA} (6.890~{\AA}), $d_1=1.389$~{\AA}
(1.408~{\AA}), $d_2=1.458$~{\AA} (1.426~{\AA}), and $d_t=1.232$~{\AA}
(1.223~{\AA}). Our determined structural parameters are in good
agreement with other studies.~\cite{Kim2012,Perkgoz2014}

To compare the structural stability of three types of graphyne, we
evaluated their cohesive energies with respect to that of graphene as
a function of $N_{sp}/N$, where $N_{sp}$ and $N$ are the number of
$sp$-bonded or doubly-coordinated atoms, and $N$ and the total number
of atoms in the unit cell, respectively. As displayed in
Fig.~\ref{structure}(d). these types of graphyne are only
$0.6-0.8$~eV/atom less stable than graphene. This tendency agrees
well with quantum Monte Carlo calculations~\cite{Ivanovskii2013,
Yin2013,Shin2014}. Especially, the $\gamma$-graphyne and
$\beta$-graphyne are even more stable than graphdiyne, which was
already reported to be synthesized,~\cite{Li2010,Qian2012} expecting
to be synthesized as well.

\begin{figure}[t]
\includegraphics[width=1.0\columnwidth]{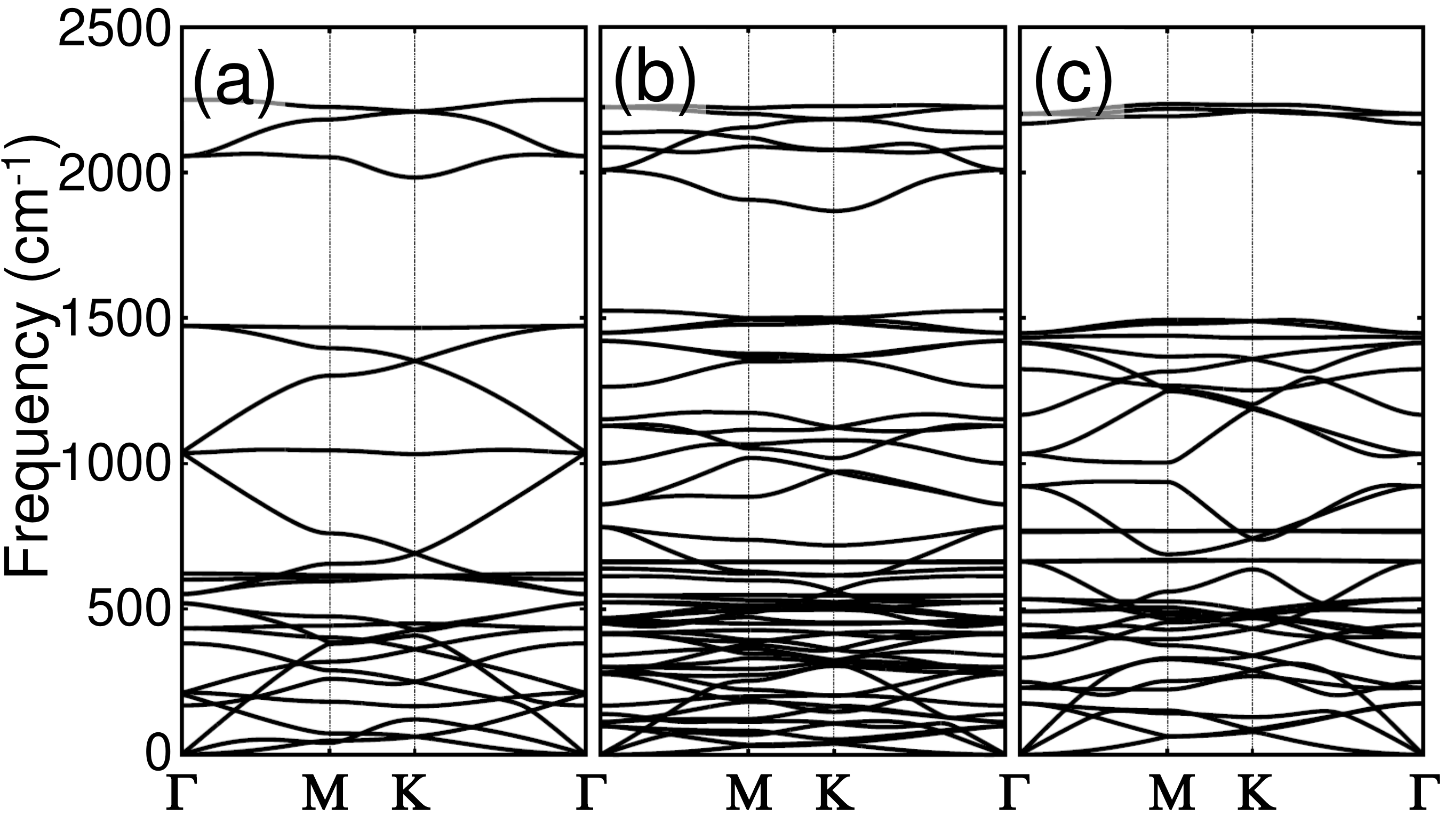}
\caption{The phonon dispersion relations of (a) $\alpha$-graphyne, (b)
$\beta$-graphyne, and (c) $\gamma$-graphyne along the high symmetry
lines in 2D hexagonal Brillouin zones.
\label{phonon}}
\end{figure}

Using $3\times3\times1$-repeated supercells of the equilibrium
structures, we calculated the phonon dispersion relations of all three
types of graphyne as shown in Fig.~\ref{phonon}, and identified all
the phonon branches by analyzing their corresponding eigenvectors. We
found in all cases that there are three acoustic phonon branches below
$\nu\approx400$~cm$^{-1}$, two in-plane modes with a linear dispersion
and one quadratic out-of-plane mode near $\Gamma$ point. We did not
find any imaginary frequencies in all the types of graphyne indicating
their structural stability. Note that there were inconsistent results
for $\beta$-graphyne in previous studies showing imaginary phonon
frequencies~\cite{Perkgoz2014} or phonon mode
anomalies,~\cite{Popov2013} which were not observed in our
calculation. We believe that such inconsistency was due to use of an
incompletely relaxed structure in a small supercell size. On the other
hand, the calculated phonon dispersion relations of $\alpha$-graphyne
and $\gamma$-graphyne are in good agreement with other
studies.~\cite{Popov2013,Perkgoz2014}

Compared to graphene with its maximum frequency of
${\lesssim}1600$~cm$^{-1}$,~\cite{Mounet2005} graphyne exhibits even 
higher phonon branches dispersed at $\nu\gtrsim2000$~cm$^{-1}$. These
high frequency modes correspond mainly to the stretching vibration of
the triple bonds $d_t$ with larger force constants than
single-bond-like $d_1$ and $d_2$, which generate graphene-like phonon
modes at $\nu\lesssim1500$~cm$^{-1}$ below a large frequency gap of
$\Delta\nu\approx500$~cm$^{-1}$, as displayed in Fig.~\ref{phonon}.
The existence of such a large frequency gap implies that the vibration
of $d_t$ bond stretching modes is strongly protected from being
scattered by the graphene-like modes by the energy conservation
law.~\cite{Jiang2014} Another phonon gap of $\Delta\nu\approx32$, 78,
or 67~cm$^{-1}$, was also observed near $\nu\approx500$, 750, or
1000~cm$^{-1}$ for $\alpha$-, $\beta$-, or $\gamma$-graphyne,
respectively. We found that the $d_1$ and $d_2$ stretching modes are
dispersed above the gap, while all the bending modes including optical
flexural modes are settled below. Note there is an exception in the
$\alpha$-graphyne as displayed in Fig.~\ref{phonon}(a) that there are
two flat modes identified as optical flexural modes located at
$\nu\approx602$ and 622~cm$^{-1}$ above the gap. Therefore, in the
$\alpha$-graphyne, these optical flexural phonons can be scattered by
the stretching phonons, while in the other types of graphyne, they are
located at similar frequencies, but below their frequency gap, as
shown in Fig.~\ref{phonon}(b) and (c), which protects them from being
scattered by their corresponding $d_1$ and $d_2$ stretching modes.

\begin{figure*}[t]
\includegraphics[width=1.0\textwidth]{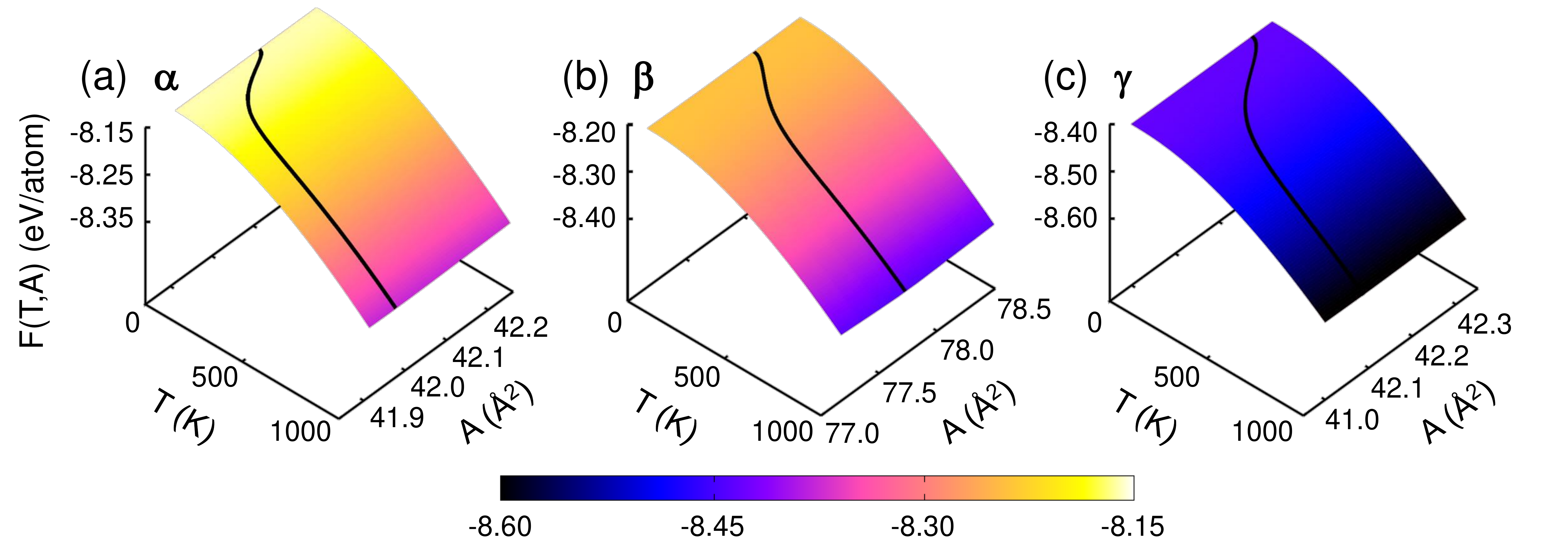}
\caption{(Color online) Color-coded maps of Helmholtz free energy of
(a) $\alpha$-graphyne, (b) $\beta$-graphyne, and (c) $\gamma$-graphyne
as a function of temperature and area calculated using
Eq.~\ref{equation1}. The black solid line on each plot indicates
minimum value of free energy at each given temperature with respect to
the area, and thus the equilibrium area for a given temperature. The
color-coded free energy values are given in the color bar below.
\label{helmholtz}}
\end{figure*}

To evaluate the Helmholtz free energy, we used Eq.~(\ref{equation1})
with the phonon dispersion relations $\omega_{\mathbf{q},n}$
calculated at 12 different area points, based on QHA. For each type of
graphyne, the calculated Helmholtz free energy was plotted as a
function of temperature and area in a color-coded map as displayed in
Fig.~\ref{helmholtz}. Note that color-coded energy values clearly
represent that the $\gamma$-graphyne is more stable than the other two
types. The temperature dependence of the equilibrium area $A(T)$ was
obtained by minimizing the thermodynamic functions with respect to the
area as denoted with the black solid line on each color-coded map in
Fig.~\ref{helmholtz}. 

\begin{figure}
\includegraphics[width=0.9\columnwidth]{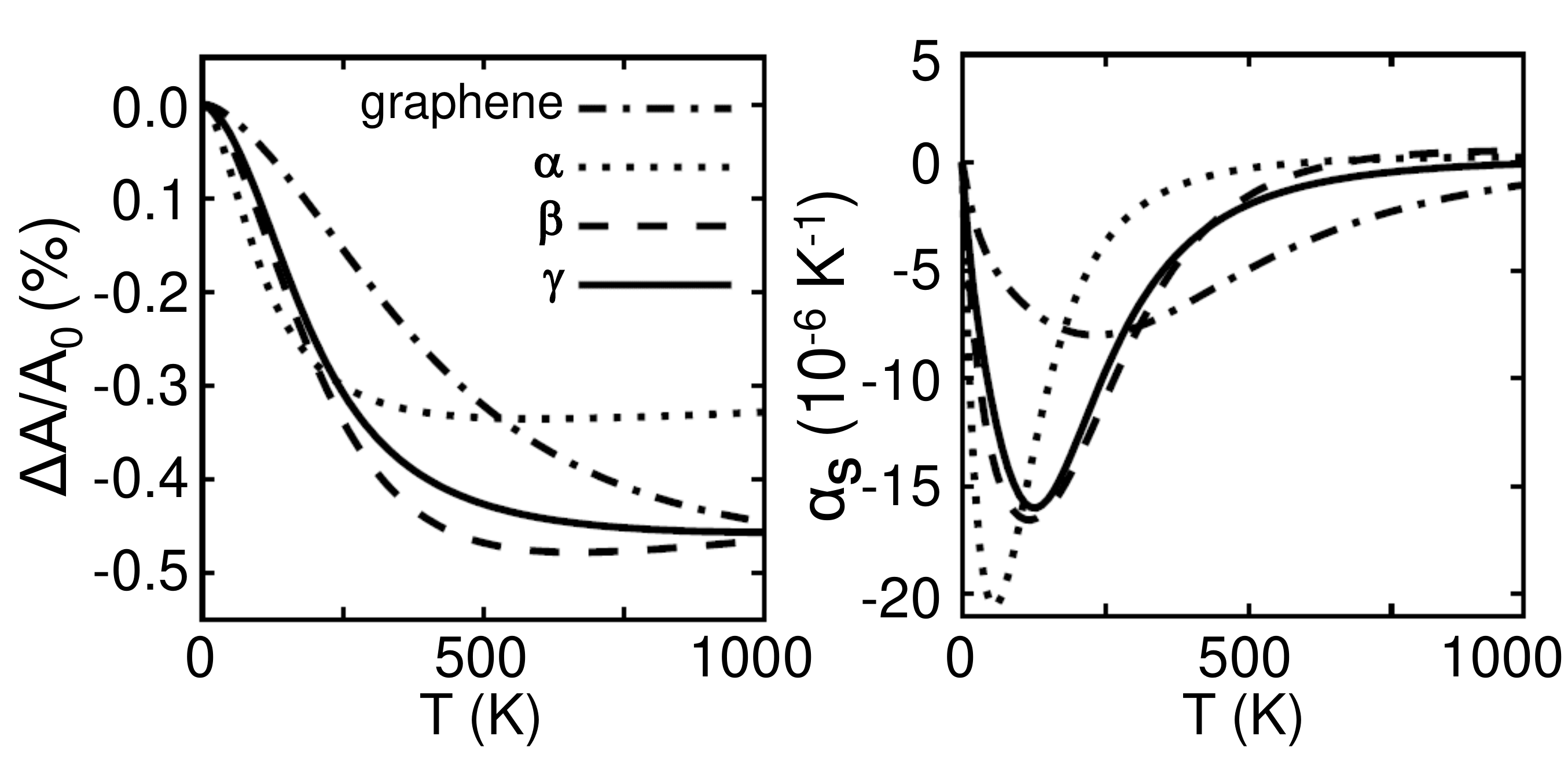}
\caption{(a) Temperature dependence of the ratio of area expansion to
the reference area at $T=0$~K, ${\Delta}A/A_0$ in the unit of
percentage \%, and (b) the corresponding area thermal expansion
coefficient, $\alpha_S(T)$, defined by Eq.~\ref{equation2}, in the
unit of $10^{-6}$~K$^{-1}$, of $\alpha$-graphyne (dotted line),
$\beta$-graphyne (dashed line), and $\gamma$-graphyne (solid line) as
well as of graphene (dash-dotted line) for comparison.
\label{TEC}}
\end{figure}

We explored the TE behaviors of three types of graphyne as well as
graphene for comparison by evaluating the temperature dependence of
change in each system's area with respect to the zero-temperature
area, ${\Delta}A/A_0\equiv[A(T)-A_0]/A_0$, and its area TEC
$\alpha_s(T)$ given by Eq.~(\ref{equation2}). It was found that
similar to graphene, all three types of graphyne contract as
temperature increases up to $T\approx1000$~K, as shown in
Fig.~\ref{TEC}(a). Such an area shrink observed in a 2D sheet was
attributed to its flexural modes, which are mainly
responsible for the ripple effect,~\cite{Lifshitz1952} as in
graphene,~\cite{Shao2012,Mounet2005} and in
graphyne.~\cite{Shao2012,Perkgoz2014} We confirmed the contribution of
flexural modes of graphyne to NTE by evaluating their corresponding
mode Gr\"uneisen parameters, $\gamma_{\mathbf{q},n}$, defined as
\begin{equation}
  \gamma_{\mathbf{q},n}=-\frac{\partial{\log{\omega_{\mathbf{q},n}}}}
     {\partial{\log{V}}}=-\frac{V}{\omega_{\mathbf{q},n}}
     \frac{\partial{\omega_{\mathbf{q},n}}}{\partial{V}},
\end{equation}
to be negative. 

Using Eq.~(\ref{equation2}), we also calculated the TECs of three
types of graphyne as well as graphene. Compared to our calculated TEC
values of graphene, which are in good agreement with previous
studies,~\cite{Shao2012,Mounet2005} it was observed that graphyne
exhibits significantly lower TEC values than graphene as shown in
Fig.~\ref{TEC}(b), which cannot be fully understood only by the
flexural modes responsible for the ripple effect. To scrutinize such
large differences, we thoroughly inspected the real-space vibrations
of all the phonon modes by visualizing their corresponding
eigenvectors together with mode Gr\"uneisen parameters at different
wave vectors $\mathbf{q}$ and mode indices $n$, and found intriguing
unexpected vibrational modes, which we classified as
RUMs.~\cite{Hammonds1996} It has already been known that the RUMs and
quasi RUMs (qRUMs) involving a little distortion of rigid polyhedra
are major causes of NTE occurred in an oxide framework consisting of
rigid polyhedra, such as MO$_4$ and MO$_6$, where M and O are a metal
cation and oxygen,~\cite{Pryde1996,Miller2009} although it was
reported that there is no simple correlation between NTE and RUMs.~\cite{Tao2003}

\begin{figure}
\includegraphics[width=1.0\columnwidth]{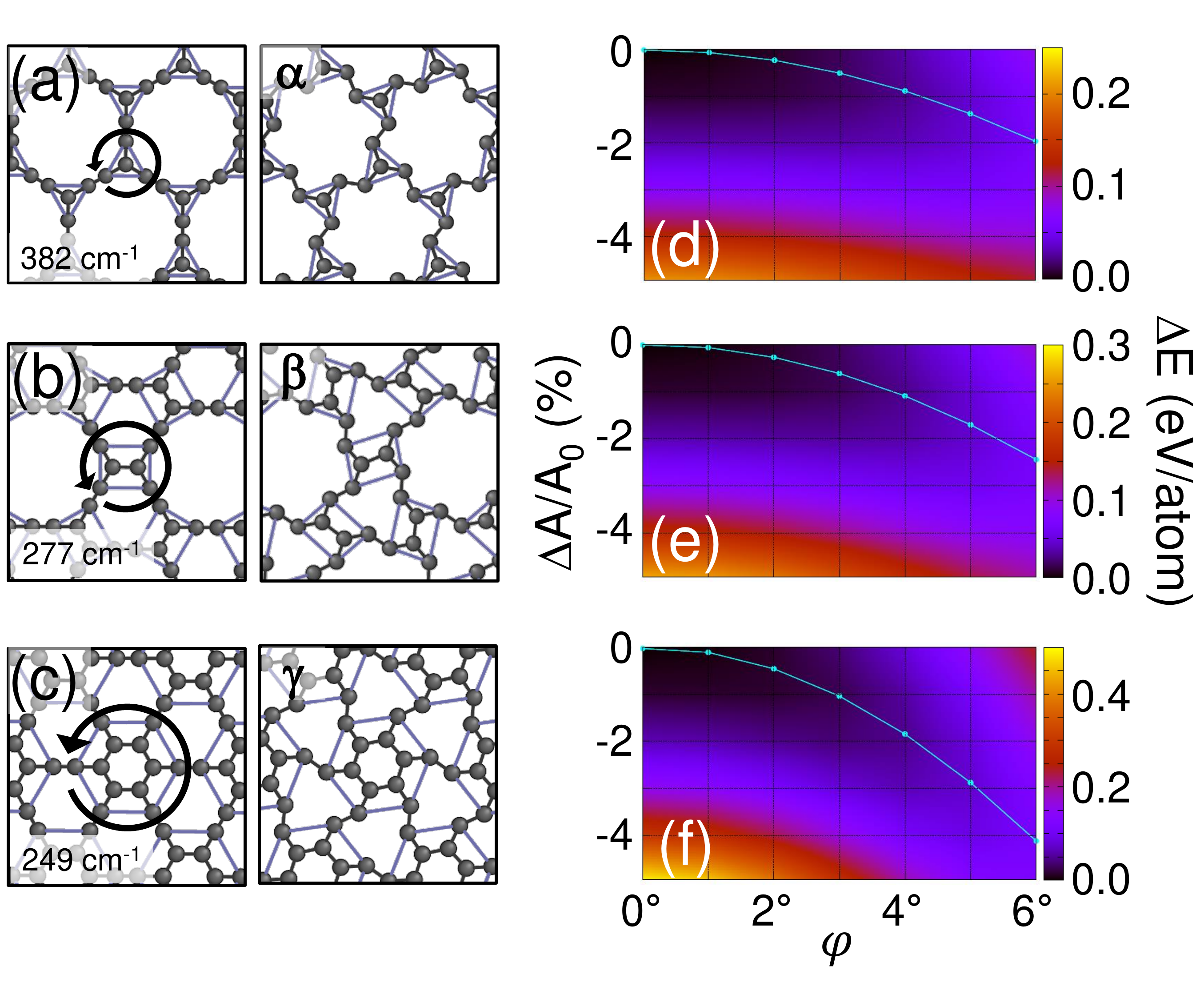}
\caption{Real-space visualization of typical rigid unit modes
observed in (a) $\alpha$-graphyne (b) $\beta$-graphyne, and (c)
$\gamma$-graphyne respectively with $\nu\approx382$, 277 and
249~cm$^{-1}$. Each local rigid unit is represented by gray polygon
(triangle, square, or hexagon in (a), (b), or (c), respectively)
surrounding $sp^2$ carbon atoms with $d_1$ and $d_2$ bonds. Each
equilibrium structure is depicted in the left box, while the right box
shows a snapshot of rigid units rotated counterclockwise
corresponding to the selected RUM. Relative energy map color-coded as
a function of ${\Delta}A/A_0$ and rotation angle $\varphi$ of each
rigid unit calculated for (d) $\alpha$-graphyne, (e) $\beta$-graphyne,
and (f) $\gamma$-graphyne. Cyan-colored solid line with dots indicates
the area contraction determined by the minimum energy for given
$\varphi$. Energy values are given in terms of the equilibrium energy
of each graphyne.
\label{RUM}}
\end{figure}

Fig.~\ref{RUM}(a--c) shows real-space visualization of typical RUMs
(in the right) observed at $\nu\approx382$, 277, and 249~cm$^{-1}$ in
$\alpha$-, $\beta$-, and $\gamma$-graphyne, respectively, in tandem
with their corresponding equilibrium configurations (in the left). The
corresponding rigid units are denoted respectively by gray polygons of
a triangle, a rectangle, and a hexagon, each of which encloses the
$sp^2$ carbon atoms connected with $d_1$ and $d_2$ bonds. During the
libration mode of the rigid unit in each graphyne, every AL becomes
bended and thus the $d_t$ bonds are lengthened as shown in
Fig.~\ref{RUM}(a--c). We found that such lengthened bonds can be
compensated by filling an empty space or reducing the area size. To
clarify the correlation between these RUMs and NTE in the graphyne
systems and to analyze the size reduction quantitatively, we
calculated the relative energy of each type of graphyne as a function
of the area change ${\Delta}A/A_0$ with respect to the
zero-temperature area $A_0$, and the rotation angle $\varphi$ of each
rigid unit. Fig.~\ref{RUM}(d--f) displays the color-coded energy
map, in which the area reduction was estimated by searching for the
minimum values of the relative energy ${\Delta}E$ for a given
$\varphi$, as depicted with cyan-colored line with dots. Since the
maximum rotation angle may increase with temperature, RUMs as well as
the ripple effects may contribute to the area reduction. The mode
Gr\"uneisen parameters corresponding the RUMs was also calculated to
be negative as similarly in the flexural modes contributing to the
ripple effects.

\section{Conclusions}
\label{Summary}

We presented the phonon dispersion relations and the thermal expansion
behaviors of $\alpha$-, $\beta$- and $\gamma$-graphyne using the
density functional theory. Our calculated phonon dispersion relations
showed that there were no imaginary frequencies implying their
structural stability. Their high-frequency modes at
$\nu\approx2000$~cm$^{-1}$ correspond to the stretching modes of the
triple bonds $d_t$, which are protected by the large frequency gap
from being scattered by graphene-like modes of $d_1$ and $d_2$ bonds.
Similarly, the stretching modes of $d_1$ and $d_2$ bonds can also be
protected from being scattered by their bending modes, since the
latter modes are separately located below another frequency gap, with
an exception in $\alpha$-graphyne. Our quasi harmonic approximation
calculations revealed that all types of graphyne exhibit negative
thermal expansion as in other 2D materials, but they have much larger
values in negative thermal expansion coefficients than graphene. This
significant discrepancy was resolved by identifying the rigid unit
modes, which were firstly observed in 2D carbon graphitic structures.
We found that not only the ripple effect originating from the flexural
modes, but also the rigid unit modes are strongly responsible for the
negative thermal expansion in graphyne.

\section*{Acknowledgments}

We thank Mr. Hyeonsu Lee for fruitful discussion. We acknowledge
financial support from the Korean government through National Research
Foundation (NRF-2015R1A2A2A01006204). Some portion of our
computational work was done using the resources of the KISTI
Supercomputing Center (KSC-2014-C2-028 and KSC-2014-G2-001).


\end{document}